\documentclass[preprint,prd,noshowpacs]{revtex4}
\usepackage{amssymb}
\usepackage{amsmath}

\begin{document}

\title{Hawking radiation, Unruh radiation and the equivalence principle}

\author{Douglas Singleton}
\email{dougs@csufresno.edu}
\author{Steve Wilburn}
\email{swilburn@csufresno.edu}
\affiliation{Physics Department, CSU Fresno, Fresno, CA 93740 USA}

\date{\today}

\begin{abstract}
We compare the response function of an Unruh-DeWitt detector for different space-times and
different vacua and show that there is a {\it detailed} violation of the equivalence principle.
In particular comparing the response of an accelerating detector to a detector at rest in a
Schwarzschild space-time we find that both detectors register thermal radiation, but
for a given, equivalent acceleration the fixed detector in the Schwarzschild space-time measures a higher temperature. This allows
one to locally distinguish the two cases. As one approaches the horizon the two temperatures have
the same limit so that the equivalence principle is restored at the horizon.
\\
\end{abstract}

\maketitle

{\bf Introduction:} The equivalence principle is the conceptual basis for general relativity \cite{ep}. It equates a gravitational field with a
uniformly accelerating reference frame {\it locally} -- by making measurements in a small enough region of space-time
one can not distinguish a gravitational field from a uniformly accelerating frame of reference. ``Small enough" means that
one does not notice the tidal forces of the gravitational field. Here we investigate the Hawking and Unruh radiation detected (or not detected)
by a detector in Schwarzschild space-time versus  Rindler space-time. Whether radiation is detected or not
depends not only on the space-time but also on the vacuum state -- Boulware vacuum, Unruh vacuum, Minkowski vacuum, Hartle-Hawking vacuum or Rindler vacuum.
We find that a detector in Rindler space-time with respect to Rindler vacuum is comparable to a detector in Schwarzschild space-time with respect to
Boulware vacuum. In this case both detectors do not detect radiation and the equivalence principle is valid.
Next we find that a detector in Rindler space-time with respect to Minkowski vacuum is comparable to a detector in Schwarzschild space-time with respect to
Unruh vacuum. In this case both detectors detect radiation, {\it but} for a given, equivalent acceleration the detector in the Schwarzschild space-time
measures a higher temperature. This gives a detailed violation of the equivalence principle.
As the detector in the Schwarzschild space-time approaches the horizon the two temperatures approach the same value.
Thus near the horizon the equivalence principle is restored.

Before proceeding to the calculations one can ask what is the conceptual basis for this violation
of the equivalence principle in the situations described above. The reason rests with the local nature of the equivalence principle
versus the non-local nature of quantum phenomena. 
If one is allowed to make global space-time measurements then one can distinguish between a uniform 
acceleration and a gravitational field. On the other hand quantum mechanics has some inherent non-locality. The prime
example is Bell's inequality experiments where particles have a non-local entanglement with one another. Also in quantum field
theory one expands a field in terms of normal modes. In flat space these modes are non-local plane waves.
Thus it is not surprising that a quantum effect like Hawking radiation or Unruh radiation should violate the equivalence principle.
The surprise is that the equivalence principle is restored as the gravitational field becomes more intense i.e. at the event horizon.
Finally, we note that there have been other suggestions that the quantum phenomenon of neutrino oscillations \cite{gasperini} \cite{ahluwalia}
\cite{mureika} violate the equivalence principle. 

{\bf Unruh-Dewitt detector in various space-times and vacua:}
To determine if an observer measures radiation we use the standard Unruh-Dewitt detector coupled to a massless
scalar field, $\phi (x)$ which has two energy levels $E_0 < E$.
References \cite{BD} \cite{akhmedov} have details of the construction of this type of detector.  
The detector-field coupling is given by the interaction $g \mu (\tau) \phi( x (\tau))$
with $g$ being the coupling constant, $\mu (\tau)$ is the detector's monopole moment, and $x(\tau)=x^\mu (\tau)$ is
the detector's trajectory as a function of its proper time, $\tau$.  The transition rate per unit proper time, $T(E)$, for such
a detector to be excited from its ground state $E_0$ to a higher energy $E$ is given by (here and throughout the paper we set $G=c=1$)
\begin{equation}
\label{t-rate}
T (E) = g^2 \sum _E | \langle E | \mu (0) | E_0 \rangle | ^2 \int _{-\infty} ^{+ \infty} e^{-i(E-E_0) \Delta \tau} G^+ (\Delta \tau) d(\Delta \tau )
\end{equation}

In the above expression $G^+ (x , x')= \langle 0 | \phi (x) \phi (x') | 0 \rangle$ is the positive frequency Wightman function
since we are studying excitations from $E_0$ to $E$ and $\Delta \tau = \tau - \tau '$. For us the important part of 
\eqref{t-rate} it the  response function {\it per unit proper time} \cite{BD}
\begin{equation}
\label{response}
{\cal F} (E)=\int _{-\infty} ^{+ \infty} e^{-i(E-E_0) \Delta \tau} G^+ (\Delta \tau) d (\Delta \tau).
\end{equation}
It is important
to note that $G^+$ is defined with respect to some vacuum state $|0 \rangle$ and picking a different vacuum state can lead to
different ${\cal F} (E)$'s. Some vacuum choices can lead to the detector being excited while other choices of vacuum leave the
detector in the ground state. It is this subtle issue of the choice of vacuum that prevents simple violations of the equivalence principle.
${\cal F} (E)$ depends of the space-time trajectory of the detector but is independent of its coupling and monopole moment. 
To simplify the calculations we take space-time as 2D. We do not
lose any essential features of the response function in this way. We will write the metrics for
our 2D space-times in light front coordinate form
$ds^2 = C(u_i,v_i) du_i ~dv_i$.
The index $i$ will indicate the particular space-time {\it and} vacuum being considered.
Different forms of the light front form metric give different forms of the wave equation with different
normal mode solutions -- $e^{-i \omega u_i}$ and $e^{-i \omega v_i}$ where $\omega$ is the energy of the mode.
From \cite{BD} the 2D Wightman function for metrics in the light front form is
\begin{equation}
\label{wightman}
G^+ (x, t; x', t') = -\frac{1}{4 \pi} \ln \left[(\Delta u_i - i \epsilon) (\Delta v_i - i \epsilon) \right] ~,
\end{equation}
where $\Delta u_i = u_i (x,t) - u_i (x',t')$ and $\Delta v_i = v_i (x,t) - v_i (x',t')$.
The crucial issue -- which vacuum the response function is being calculated with respect to -- is embedded in the
form of the 2D Wightman function in \eqref{wightman}. This in turn is determined by the specific form of the metric. 
The character of the vacuum is determined by the form of the
metric in the following way: (i) The form of the metric determines the specific form of the wave equation in the space-time
which in turn determines the normal mode solutions, $u_k (x,t)$. (ii) A field, $\phi$, can be expanded in terms of these
modes as $\phi = \sum _k (a_k u_k + a_k ^{\dagger} u^* _k)$. (iii) Turning the $a_k , a^{\dagger} _k$ into annihilation
and creation operators then defines the vacuum e.g. $a_k | 0 \rangle =0$ ; $a^{\dagger} _k | 0 \rangle =|1 \rangle$.

{\underline {\it Minkowski space-time:}} Since for two of the cases studied below -- Schwarzschild space-time for Boulware vacuum
and Rindler space-time for Rindler vacuum -- the response function reduces to essentially Minkowski space-time
we will study this example first. The light front coordinates for Minkowski space-time are $u_{MM}=t-x$ , $v_{MM}=t+x$,
the conformal factor is $C(u,v)=1$. The subscript $i=MM$ stands for Minkowski space-time and Minkowski
vacuum. Taken together this gives $ds^2=dt^2-dx^2 = du_{MM} dv_{MM}$.
For a detector at rest or moving with uniform velocity, $v$, in Minkowski space-time one
has $\Delta u_{MM} = \Delta v_{MM} = \Delta \tau$. If the detector is at rest $\Delta \tau = \Delta t$ while if the
detector is moving $\Delta \tau = \Delta t \sqrt{1-v^2}$. From \eqref{wightman} this Wightman function is
\begin{equation}
\label{wightman-min}
G^+ _{MM} = -\frac{1}{4 \pi} \ln \left[ (\Delta \tau - i \epsilon) ^2 \right] ~.
\end{equation}
(For a detector moving with uniform velocity $v$ one must absorb a factor of $1/\sqrt{1-v^2}$ into
$\epsilon$). Inserting \eqref{wightman-min} into the response function \eqref{response} and evaluating
the integral via a contour integration gives ${\cal F}_{MM} (E) =0$ since $E > E_0$. This is what one expects --
an inertial detector in Minkowski will not spontaneously get excited.

{\underline {\it Rindler space-time:}} Next we turn to Rindler space-time which is Minkowski space-time as seen by a uniformly accelerated observer
with acceleration $a$. The path of such an observer is given by
\begin{equation}
\label{rindler}
t= \frac{1}{a} \sinh(a \tau) ~~; ~~ x= \frac{1}{a} \cosh(a \tau) ~,
\end{equation}
where $\tau$ is the detector's proper time. The Rindler space-time in light front form is
\begin{equation}
\label{rindlera}
ds^2 = dt^2- dx^2 = du_{RM} dv_{RM} ~,
\end{equation}
where $u_{RM} = t - x$ , $v_{RM} = t + x$ giving $(\Delta u_{RM} - i \epsilon)(\Delta v_{RM} - i \epsilon) =
(\Delta t - i \epsilon )^2 - (\Delta x)^2$. The coordinates $t$ and $x$ are given by \eqref{rindler}.
The subscript $RM$ stands for Rindler space-time and Minkowski vacuum state.
This is explained in more detail below. Using \eqref{rindler} this becomes
$$
\frac{1}{a^2} \left[ (\sinh(a \tau) - \sinh (a \tau') - i \epsilon )^2 - ((\cosh(a \tau) - \cosh(a \tau'))^2 \right] 
= \frac{4}{a^2} \sinh ^2 \left( \frac{a (\Delta \tau - i \epsilon )}{2} \right)
$$
Using these results in \eqref{wightman} gives the Wightman function for this form of Rindler
\begin{equation}
\label{wightman-rindler}
G^+ _{RM} = -\frac{1}{4 \pi} \ln \left[ \frac{4}{a^2} \sinh ^2 \left( \frac{a (\Delta \tau - i \epsilon )}{2} \right)\right]
\end{equation}
Inserting this Wightman function into the response function \eqref{response}, and performing a contour integration
gives \cite{BD} \cite{brout} a Planckian response function
\begin{equation}
\label{t-unruh}
{\cal F} _{RM} (E) \propto \frac{1}{E( e^{E/k_B T_{RM}} -1)} ~~~~ {\text{where}} ~~~~ k_B T_{RM} = \frac{a}{2 \pi}.
\end{equation}
where $k_B$ is Boltzmann's constant.
This is the Unruh temperature given in terms of the acceleration, $a$, of the observer. The vacuum state
associated with the form of the Rindler metric given by \eqref{rindlera} is called the
Minkowski vacuum \cite{ginzburg} \cite{grishchuk}. It is with respect to this vacuum state that an
observer will have a non-zero response function and will detect particles.

Rindler space-time can also be cast in Rindler coordinates $(\eta, \zeta)$ which are defined via
$t= \frac{e^{a \zeta}}{a} \sinh(a \eta) ~~; ~~ x= \frac{e^{a \zeta}}{a} \cosh(a \eta)~$,
In these coordinates the Rindler metric is
\begin{equation}
\label{rindler2a}
ds^2 = e^{2 a \zeta} (d \eta ^2 - d \zeta ^2) = e^{2 a \zeta} du_{RR} dv_{RR} ~,
\end{equation}
where the final form is in terms of light front coordinates $u_{RR} = \eta - \zeta$ and
$v_{RR} = \eta + \zeta$. The subscript $RR$ stands for Rindler space-time with respect the
the Rindler vacuum state. The coordinate $\eta$ plays the role of time and $\zeta$ plays the role of
position. The proper time of the detector is $\tau = e^{a \zeta} \eta$ and
$a e^{- a \zeta}$ is the proper acceleration. In this set of coordinates different
$\zeta$'s correspond to picking different proper accelerations. Thus for a detector with
a fixed proper acceleration $\zeta = const.$ so that $\Delta u_{RR} = \Delta v_{RR} =
\Delta \eta = e^{-a \zeta} \Delta \tau$. The resulting Wightman function is then the
same as that for Minkowski space-time given in \eqref{wightman-min}. One must now absorb a
factor of $e^{a \zeta}$ into the $i \epsilon$ term. Thus as for Minkowski space-time the
response function is zero, ${\cal F}_{RR} (E) =0$. This vacuum Rindler vacuum state
has been used \cite{ginzburg} to preserve the equivalence principle
against the following argument : ``If a detector in uniform accelerated motion detects radiation
but a detector at rest in a gravitational field without a horizon (e.g. a detector in the gravitational field
of the Earth) does not detect radiation can't one in this way distinguish a gravitational field
from an acceleration?" The answer is that one needs to compare the two cases in question
in the appropriate vacuum state -- what Ginzburg and Frolov \cite{ginzburg} call ``corresponding vacua". Thus one should
compare the Rindler vacuum of the accelerated observer with the Boulware vacuum (this vacuum state is discussed next)
of the detector fixed in a gravitational field without a horizon. In this way the equivalence principle is preserved against
the preceding argument.

{\underline {\it Schwarzschild space-time:}} We now calculate the response function for detector in Schwarzschild space-time.
Starting from the form of the metric $ds^2 = (1-2M/r) dt^2 - (1-2M/r)^{-1} dr^2$ we
transform this into the light front form 
\begin{equation}
\label{sch1}
ds^2 = \left( 1-\frac{2M}{r} \right) du_{SB} ~dv_{SB} ~.
\end{equation}
In this equation $u_{SB}=t-r^*$ and $v_{SB}=t+r^*$ with
$r^* = \int \frac{dr}{1-2M/r} = r + 2M \ln |r/2M- 1 |~$.
The subscript $SB$ stands for Schwarzschild space-time in the Boulware vacuum state.
For a detector at a fixed radius $r=R$,  $r^* \rightarrow R^* = R + 2M \ln (R-2M)$ and so $\Delta r^* =0$. 
Thus $\Delta u_{SB} =\Delta v_{SB} = \Delta t$. The relationship between proper time, $\tau$, and Schwarzschild time, $t$, is
$\Delta \tau = \sqrt{1 - \frac{2M}{R}} \Delta t $.
Combining these results we see that the Wightman function for the Schwarzschild metric of the form \eqref{sch1} is essentially
the same as the Minkowski space-time Wightman function in \eqref{wightman-min}. One must absorb the constant factor $\sqrt{1-2M/R}$
into $\epsilon$. Thus as for Minkowski space-time the response function for this form of the
Schwarzschild metric is zero, ${\cal F}_{SB} (E) =0$. This result may seem surprising since it appears we have shown Hawking radiation
does not exist. Actually what this shows is that there is no radiation with respect to the vacuum state defined
by the choice of the Schwarzschild metric in \eqref{sch1}. This vacuum state
\eqref{sch1} is called the Boulware vacuum \cite{boulware}. There is no Hawking radiation with respect
to the Boulware vacuum. However the Boulware vacuum is not physical near the horizon where the energy momentum tensor
diverges.

Two vacuum states which are well behaved at the horizon are the Hartle-Hawking vacuum \cite{hartle} and Unruh vacuum \cite{unruh}. 
To study these two vacuum states we write the Schwarzschild metric in Kruskal form
$$
ds^2 = \frac{2M}{r} e^{-r/2M} du_{SH} dv_{SH} ~,
$$
with $u_{SH} = -4M e^{-u_{SB}/4M}$ , $v_{SH} = 4M e^{v_{SB}/4M}$ where $u_{SB}, v_{SB}$ were previously defined and $r =r(u_{SH}, v_{SH})$ is
an implicitly defined function of $u_{SH}, v_{SH}$. The subscript $SH$ stands for the Schwarzschild space-time in the
Hartle-Hawking vacuum. The Hartle-Hawking vacuum uses the coordinates, $u_{SH} , v_{SH}$, to calculated $\Delta u_{SH}, \Delta v_{SH}$ which
are then used to calculate the associated Wightman function. The Hartle-Hawking vacuum corresponds to the state when a Schwarzschild 
black hole is in equilibrium with a thermal bath which is at the same temperature as that of the black hole. For a realistic black hole formed via 
collapse the Unruh vaccum is a better choice. The Unruh vacuum corresponds to the Boulware vacuum in the far past and the Hartle-Hawking vacuum 
in the far future. In detail for the Unruh vacuum a detector fixed $r=R$ has $\Delta u_{SU}, \Delta v_{SU}$ given by \cite{BD}
\begin{equation}
\label{d-unruh}
\Delta u_{SU} = \Delta u_{SH} = - 4 M e^{R^*/4M} (e^{-t/4M} - e^{-t'/4M}) ~~~, ~~ \Delta v_{SU} = \Delta v_{SB} = \Delta t ~.
\end{equation}
$R^*$ is defined by setting $r=R$ in the expression for $r^*$; $\Delta t = t -t'$. Using \eqref{d-unruh} gives
\begin{eqnarray}
\label{uv-unruh}
&&(\Delta u_{SU} - i \epsilon)(\Delta v_{SU} -i \epsilon) = 8 M e^{R^*/4M}  
e ^{-(t+t')/8M} (\Delta t - i\epsilon ) \sinh \left( \frac{\Delta t -i \epsilon}{8M} \right) \nonumber \\
 &=& 8 M e^{R^*/4M} e ^{-(t+t')/8M} \left( \frac{\Delta \tau - i \epsilon}{\sqrt{1 - 2M/R}} \right) 
\sinh \left( \frac{\Delta \tau - i \epsilon}{8 M \sqrt{1 - 2M/R}} \right) ~.
\end{eqnarray}
Inserting this result into \eqref{wightman} gives the Wightman function which is
essentially the same as the Wightman function for the Rindler metric in the Minkowski vacuum \eqref{wightman-min} but with
$a$ replaced by $1/(4 M \sqrt{1 - 2M/R})$, which is the blue shifted surface gravity $\kappa = 1/4 M$ of the 
black hole. Inserting this Wightman function in the response function \eqref{response} then leads to the same type of contour integral as
before. The factor $e ^{-(t+t')/8M}$ does not contribute to the contour integration and the multiplicative factor, $\Delta \tau - i \epsilon$, 
is essentially that for Minkowski space-time and also does not contribute. The result is again a Planckian spectrum
\begin{equation}
\label{t-hawking}
{\cal F}_{SU} (E) \propto \frac{1}{E( e^{E/k_B T_{SU}} -1)} ~~~~ {\text{where}} ~~~~ k_B T_{SU} = \frac{1}{8 \pi M \sqrt{1 - 2M/R}}
\end{equation}
which is the temperature measured by the fixed detector with respect to the Unruh vacuum for Schwarzschild space-time.
Doing the calculation in the Hartle-Hawking vacuum yields the same temperature. The only difference is that
the equation \eqref{uv-unruh} for the Hartle-Hawking vacuum is proportional to $\sinh ^2 (...)$ rather than
$\sinh (...)$. This leads to an unimportant, multiplicative factor of two in ${\cal F}_{SH} (E)$. This result for
$T_{SU}$ from \eqref{t-hawking} is consistent with the higher dimensional embedding approach of \cite{deser} \cite{thorlacius}.
In these works the temperature of the static Schwarzschild observer in \eqref{t-hawking} is obtained as the Unruh
temperature in a six dimensional space-time  using an effective six-acceleration equals the blue shifted surface gravity,
$\kappa = 1/(4 M \sqrt{1 - 2M/R})$.  
The magnitude of the real acceleration measured by this fixed observer, $r=R$, in the Schwarzschild space-time, 
is given by \cite{carroll}
\begin{equation}
\label{acc-sch}
a_S = \frac{{\sqrt{\nabla _\mu V \nabla ^\mu V}}}{V} = \frac{M}{R^2 \sqrt{1-2M/R}}~,
\end{equation}
where $V=\sqrt{1-2M/r}$ is the redshift factor for the Schwarzschild space-time. A similar calculation for the Rindler space-time
(using the Schwarzschild-like form of the Rindler metric $ds^2 =(1+2 a x) dt^2 - (1+2 a x)^{-1} dx^2$) yields $a_R =a$ as expected.
It is the results in \eqref{t-hawking} and \eqref{acc-sch} which lead to a violation of the equivalence principle when compared to
similar measurements in Rindler space. Both the fixed observer in Schwarzschild and the
observer in Rindler space should measure the same acceleration so we set $a_R=a=a_S$ where $a$ is the acceleration of
the Rindler observer. If we then substitute this into \eqref{t-unruh} we get
\begin{equation}
\label{t-unruh2}
k_B T_{RM} = \frac{M}{2 \pi R^2 \sqrt{1-2M/R}} ~.
\end{equation}
By comparing \eqref{t-hawking} and \eqref{t-unruh2} it is clear that $T_{RM} < T_{SU}$ when $R>2M$. Thus one can tell the
two systems apart by making local space-time measurements. As one approaches the horizon,
$R=2M$, the two results converge to the same (infinite) value. Thus at the horizon the equivalence principle is
restored. The procedure for an observer equipped with a means of measuring local acceleration and an
Unruh-Dewitt detector for measuring temperature is as follows: (i) Measure the local acceleration. (ii) Insert this
into \eqref{t-unruh}. (iii) Measure the temperature via the Unruh-Dewitt detector. If the temperature is higher than
that calculated in step (ii) then one is in a gravitational field and not in an accelerating frame.

{\bf Summary:}
In this paper we have shown that there is a violation of the equivalence principle when, for the same locally measured acceleration, one compares the
temperature of a detector fixed at $r=R>2M$ in the background of a Schwarzschild black hole and a uniformly accelerating
detector. Both detectors will detect thermal radiation, but for equal, local accelerations the detector in the gravitational background
will measure a higher temperature than the accelerating detector. The subtle feature in the analysis arises in
that one must compare the response function of the detector with respect to what Ginzburg and Frolov \cite{ginzburg} call
``corresponding" or ``matched" vacua. When comparing an accelerating detector in the Rindler vacuum with a fixed detector in
Schwarzschild space-time in the Boulware vacuum, both detectors will not detect radiation. On the other
hand comparing an accelerating detector in Minkowski vacuum with a fixed detector in Schwarzschild space-time in Unruh vacuum or Hartle-Hawking vacuum,
both detectors will detect thermal radiation. Up to this point the equivalence principle works qualitatively. However,
if one compares the values of the temperature of the thermal radiation measured in each case -- \eqref{t-unruh} for the accelerating detector
and \eqref{t-hawking} for the detector fixed in a gravitational field -- one finds that for the same acceleration, the detector
in gravitational field will measure a higher temperature thus allowing one to tell the two cases apart. As $R \rightarrow \infty$,
$T_{SU} \rightarrow 1/8 \pi M$ while from \eqref{t-unruh2} $T_{RM} \rightarrow 0$. This latter result occurs since as
$R \rightarrow \infty$ the acceleration due to gravity goes to $0$ as $1/R^2$ i.e. neither the acceleration \eqref{acc-sch}
nor the associated temperature $T_{RM}$ from \eqref{t-unruh2} are long range. Hawking radiation, since it is a radiation field, falls off like
$1/R$ and is a long range, and thus does have a constant flux/temperature as $R \rightarrow \infty$. Conversely  
as one approaches the horizon the two temperatures, $T_{SU}$ and $T_{RM}$, approach the same (infinite) value. 
Thus quantum field theory/quantum mechanics appears to become more compatible with the equivalence principle and 
general relativity as one approaches the extreme conditions near a black hole
horizon. This could optimistically be taken as a hint that gravity and quantum mechanics become more compatible and begin to merge into a
consistent theory of quantum gravity at high energies/extreme gravitational fields.

\end{document}